\documentclass[manuscript,screen,nonacm]{acmart}
\AtBeginDocument{%
  \providecommand\BibTeX{{%
    \normalfont B\kern-0.5em{\scshape i\kern-0.25em b}\kern-0.8em\TeX}}}

\usepackage{tabularx}

\newcommand\blfootnote[1]{%
  \begingroup
  \renewcommand\thefootnote{}\footnote{#1}%
  \addtocounter{footnote}{-1}%
  \endgroup
}

\setcopyright{acmcopyright}
\copyrightyear{2020}
\acmYear{2020}
\acmDOI{10.1515/icom-2020-0025}

\acmConference[Conference acronym 'XX]{Make sure to enter the correct
  conference title from your rights confirmation emai}{June 03--05,
  2018}{Woodstock, NY}
\acmBooktitle{i-com, vol. 19, no. 3} 
\acmPrice{15.00}
\acmISBN{978-1-4503-XXXX-X/18/06}

\begin{document}

\title{Examining Autocompletion as a Basic Concept for Interaction with Generative AI}

\author{Florian Lehmann}
\email{florian.lehmann@uni-bayreuth.de}
\affiliation{%
  \institution{Research Group HCI + AI, Department of Computer Science, University of Bayreuth}
  \city{Bayreuth}
  \country{Germany}
}

\author{Daniel Buschek}
\orcid{0000-0002-0013-715X}
\email{daniel.buschek@uni-bayreuth.de}
\affiliation{%
  \institution{Research Group HCI + AI, Department of Computer Science, University of Bayreuth}
  \city{Bayreuth}
  \country{Germany}
}

\renewcommand{\shortauthors}{Lehmann and Buschek}

\blfootnote{\textbf{This is an author version of the original article:} \newline
Lehmann, Florian and Buschek, Daniel. "Examining Autocompletion as a Basic Concept for Interaction with Generative AI" 
\newline i-com, vol. 19, no. 3, 2020, pp. 251-264. \href{https://doi.org/10.1515/icom-2020-0025}{https://doi.org/10.1515/icom-2020-0025} \vfill}

\begin{abstract}
  Autocompletion is an approach that extends and continues partial user input. We propose to interpret autocompletion as a basic interaction concept in human-AI interaction. We first describe the concept of autocompletion and dissect its user interface and interaction elements, using the well-established textual autocompletion in search engines as an example. We then highlight how these elements reoccur in other application domains, such as code completion, GUI sketching, and layouting. This comparison and transfer highlights an inherent role of such intelligent systems to extend and complete user input, in particular useful for designing interactions with and for generative AI. We reflect on and discuss our conceptual analysis of autocompletion to provide inspiration and a conceptual lens on current challenges in designing for human-AI interaction.
\end{abstract}


\maketitle

\section{Introduction}

Autocompletion is a well established key feature in many applications today. It is most commonly used in search engines, such as Google, Bing, and Elasticsearch, by millions of users every day. For instance, when a user types a request into a web search, the system creates certain extended variations of the input and serves these back to the user. The autocompleted variations represent a list of search term suggestions. Then the user is free to choose from these autocompleted suggestions. The selection can be confirmed or edited further. In the context of search engines, such textual autocompletion is often called query autocompletion (QAC). In the context of software engineering, it is commonly referred to as autocomplete.

This approach is used in search engines to assist a user in formulating a proper input. Moreover, it is used in web browsers when typing a domain name, in code editors or IDEs when writing source code, and in other software tools where it is important to support the user in finding a precise input. Autocompletion can be considered a supportive technology to declare an input. As a side effect, autocompletion helps to make input faster \cite{jakobsson_autocompletion_1986}. However, the underlying key concept remains: Autocompletion continues and extends (partial) user input.

In light of this vital concept, we interpret autocompletion as a generative approach embedded in a user interface. When speaking of generative in the context of machine learning within this paper, we refer to those approaches that can be used to generate things. Methods to provide such generations can be found in the field of machine learning as well. Despite classifying input to make predictions, machine learning can also be used to generate data. Such generative models typically learn an underlying data distribution from which to sample new outputs. For example, generative machine learning approaches can complete pictures \cite{nazeri_edgeconnect_2019, wang_image_2018, yu_generative_2018} and gestures \cite{bennett_simpleflow_2011}, or extend texts \cite{brown_language_2020, radford_language_2019, vaswani_attention_2017}. Implementing generative machine learning into interactive software tools can open new possibilities. For instance, it can be used to transform digital sketches into mock-ups \cite{pandian_blackbox_2020}, create sketches from text descriptions \cite{huang_sketchforme_2019}, or to create digital wireframes from paper based sketches \cite{buschek_paper2wire_2020}. Another real world example is Kite\footnote{Kite: \url{https://www.kite.com}}, a coding assistant based on the GPT language model\footnote{Better Language Models
and Their Implications (OpenAI GPT2): \url{https://openai.com/blog/better-language-models}}. It can generate complete methods from just a signature of methods, and a short, descriptive comment. 

Although the latter examples rely on machine learning models, it is not mandatory to do so for an application to provide a generative feature. For instance, in the case of query autocompletion, it could be built on n-gram frequency statistics~\cite{mitra_query_2015}, and layout generation could be based on integer programming~\cite{dayama_grids_2020}. In particular, when analysing generative approaches from a perspective of user interaction, it is hard to distinguish whether the application has machine learning implemented or not. The user interface functions as an abstract layer and hides the technology in the background. Accordingly, in this paper we regard ``intelligence'' as the ability to generate extended and ranked output based on partial user input, regardless of how this is technically achieved.

We assume that in the future, more and more applications will incorporate intelligent features. However, recent work by \citet{yang_re-examining_2020} pointed out challenges in designing applications with human-AI interaction. These include, for instance, the challenge of envisioning interaction with AI, in understanding AI capabilities, and in crafting interactions for unpredictable output. 

Such challenges motivate us to reflect on -- and learn from -- existing interaction solutions as one approach towards informing future designs: In particular, in this paper, we revisit autocompletion as a reoccurring and reusable interaction concept for designing interaction with generative intelligent systems.

We contribute a conceptual analysis and transfer in three steps: First, we systematically analyse the underlying interaction and user interface of textual autocompletion, and extract its key conceptual elements. 

Second, we identify these elements of textual autocompletion in other domains that use generative approaches, highlighting opportunities to transfer this concept and reuse it. 
Third, we reflect on potential benefits of this transfer in the light of the challenges of designing for human-AI interaction, and point out opportunities and challenges for future work. 

\section{Related Work}

This research originated in the context of a broad literature review on topics combining HCI and AI. Within this research process we discovered similarities between autocompletion and the capabilities of generative machine learning approaches. In the following paragraphs, we summarise relevant work related to these topics. 

\subsection{Research on Autocompletion}

Autocompletion is a broad topic with different research directions. A survey by \citet{cai_survey_2016} helps to gain a first overview of the most important topics: They indicate that most papers concentrate on the technical issues rather than on the user interface or interaction. 

\subsubsection{Frontend: User Interaction}

Early HCI research on user interactions on textual autocompletion was done in 1986 by  \citet{jakobsson_autocompletion_1986}. In Jakobsson's work, autocompletion was investigated as part of a library information system. The evaluation showed that textual autocompletion is more efficient to find entries in an information system in comparison to using shortcodes and a code catalogue. Besides research on efficiency, other work concentrated on engagement with the completed suggestions, for instance, how the input technique and suggestion ranking influences the selections by the users: Work by \citet{mitra_user_2014} observed that users are more likely to engage with the autocompleted suggestions if the fingers have to travel longer between keystrokes or at word boundaries. Moreover, they showed that top ranked suggestions were preferred. A strong position bias was also found by \citet{hofmann_eye-tracking_2014}. They used eye-tracking to investigate how ranking positions affect user interaction. In their study participants focused on top-ranked suggestions regardless of whether the list war randomised or not. Others investigated how the organisation of suggestions influences user interaction \cite{boughanem_organizing_2009}. For this, they compared alphabetical ordering with categorical ordering and composite. Their findings showed group and composite organisation to improve efficiency. Additionally, they suggest how to design for different organisation strategies.

\subsubsection{Backend: Ranking, Personalisation, Modelling}
Ranking and personalisation is a major theme in research on autocompletion with search engines. Models for improving ranking suggestions are also of importance. Thus, the core of research on backend functionalities for autocompletion concentrates on algorithms. As a part of that, research introduced an indexing data structure to improve the performance of query processing \cite{bast_type_2006}. Focusing on adding context-sensitivity to algorithms, work by \citet{bar-yossef_context-sensitive_2011} introduced and evaluated methods to incorporate users' search queries for suggestions. As well, they investigated how it affects ranking. Such context-sensitivity can be interpreted as personalisation of suggestion results. Adding personalisation to algorithms, research involved user-specific and demographic features \cite{shokouhi_learning_2013}. Selective personalisation was investigated by \citet{cai_selectively_2016}. They showed that the typed prefix can indicate when it is appropriate to display personalised suggestion rankings. In another work \cite{cai_diversifying_2016}, they introduced an approach to diversify the suggestion results. They aimed to rank the intended term as high as possible while reducing redundancy in the list. For this, they evaluated a model that relies not only on current search popularity but also on within-session context. Including time-series in a model showed to further improve suggestion quality \cite{shokouhi_time-sensitive_2012}. A comparison of eleven ranking approaches can be found in work by \citet{di_santo_comparing_2015}.

User interactions offer further possibilities to model autocompletion. For example, \citet{li_two-dimensional_2014} introduced a two-dimensional click model to better explain the vertical position bias and horizontal skipping bias. Incorporating the skipping behaviour into existing models they were able to improve efficiency. Predictions on the search intent can be done based on keystrokes and clicks \cite{li_analyzing_2015}. Furthermore, interactions with apps can be used to rank suggestions in search on mobile devices \cite{zhang_towards_2016}. Instead of such active user feedback, also implicit negative feedback, for instance skipping suggestions or dwell time can be involved to model suggestions. In particular, \citet{zhang_adaqac_2015} used dwell time and position of unselected suggestions as features for implicit negative feedback to adapt the ranking of query suggestions.

\subsection{Autocompletion is not Only for Text}

Beyond text, autocompletion can be applied to an array of other domains, for instance, sketching, image editing, animation, and many more. It was integrated into a GUI-based tool to create XML \cite{lin_lotusx_2012}. \citet{bennett_simpleflow_2011} presented approaches for gestural autocompletion. They showed that autocompletion improves gestures: These were shorter, more accurate, and faster to execute. Others focused on sketches and reported on a system to autocomplete digitally sketched symbols \cite{costagliola_investigating_2013, tirkaz_sketched_2012}. Further work proposed a framework to enable autocompletion on value cells in relational tables such as spreadsheets \cite{zhang_auto-completion_2019}. In the context of code, Pythia \cite{svyatkovskiy_pythia_2019} offers code completion based on a neural net to rank method and API suggestions. Work on creative domains demonstrated that an RNN, trained on physics-based simulations, can be used to autocomplete keyframe animations \cite{zhang_data-driven_2018}. Also, \citet{hsu_autocomplete_2020} presented autocompletion for aggregate elements that can be used for 2D planes, 3D surfaces, and 3D volumes. Feedback on their approach showed that workload can be reduced and the system encouraged participants to explore more variations.
 
\subsection{Generative Machine Learning has Autocompletion Capabilities}

Within our literature research, we observed opportunities for connecting autocompletion as a concept with generative machine learning, and vice-versa. In this light we highlight some generative machine learning approaches in the following. In particular they share the capability to make partial user input more complete. One example for that is work by \citet{park_neural_2017}, using neural networks for textual autocompletion. 

In general, generative approaches in machine learning are used to generate data based on prior observations, and are thus different, for example, from classification tasks. Generation has gained increasing attention with the rise of Deep Learning:
For instance, generative approaches can be used to model language for various NLP tasks. Work by \citet{vaswani_attention_2017} introduced Transformer networks. A Transformer relies solely on self-attention, this replaced recurrent layers which were commonly used with encoder-decoder architectures in the past. The language model GPT-3 is based on a Transformer architecture \cite{brown_language_2020}, which is a further developed version of GPT-2 \cite{radford_language_2019}. GPT-3 is considered a state-of-the-art model that can fulfill different functions: For example, it can generate news articles, translations, or correct English grammar.

Such machine learning models can also be integrated into creative applications: For example, \citet{huang_sketchforme_2019} introduced a system to generate sketches from text input. Another paper introduced AI tools that can help to support UI designers \cite{pandian_blackbox_2020}. One of their tools can detect UI elements on low-fidelity sketches, which then can be transformed into a medium fidelity mock-up. Other work presented a tool to transform analog sketches into digital wireframes \cite{buschek_paper2wire_2020}. 

However, it is not always a must for intelligent, interactive applications to rely on machine learning. In a comparison within this paper, we involved, for instance, a tool that is not based on machine learning, however, it can be used to automatically solve (and thus generate/complete) layouts for user interfaces \cite{dayama_grids_2020}. In this paper, we follow a user-centred view \cite{yang_mapping_2018, yang_re-examining_2020}: In particular, we aim to take a conceptual yet concrete step on the path towards user-centred, interactive AI by examining an existing interactive concept -- autocompletion -- in the ``new'' light of generative computational capabilities.

\section{The Concept of Autocompletion}
Our analysis starts by describing autocompletion on a conceptual level. Thereafter, we go more into detail until we have formally understood what autocompletion is. For describing autocompletion, we refer to the case of textual autocompletion since this is currently its most common application.

\subsection{Overview and Delineation}
On a conceptual level, autocompletion has the role to continue, extend or complete digital content. This could be any input made by a user. More specifically, such input is processed by a system in order to generate an extended version. For example, in probabilistic terms, the model samples continuations conditioned on the user input. These different variations of the extended input are then presented to the user. The user is then able to select one of the suggestions or ignore them. Either the completion was successful, or the user decides to further specify the original intent. Autocompletion allows for close-loop interaction where the system reacts to the user and vice versa.

As a software feature, autocompletion (also: autocomplete) is used in search engines to formalise a query, in content management systems to complete category names, and on smartphones to predict the next word.

Similar software features are auto fill and auto correct \cite{banovic_limits_2019}. Auto correct is sometimes implemented together with autocompletion. This supports the user to correct the faulty input and then suggests an autocompletion based on the corrected input. This might increase the overall convenience for the user when working with textual input. 
Auto fill instead aims to complete form input. A common technique is to detect the form field identifiers and recognise past input. If there was past input on similarly named input fields, then the software will suggest to complete the form automatically. Compared to auto correct, auto fill might less often appear together with autocomplete.

\subsection{Technical Approaches}
Here we outline some technical approaches that can be used to enable a computational system for autocompletion capabilities.

\subsubsection{Approaches in Industry and Commercial Products}
In commercial products, transparency on how systems manipulate the input to complete it, is typically not offered to the end-user. Algorithms are kept secret and the user interface works as an abstract layer for the user to keep interaction simple and hide the technical functions.

Insights from a practical perspective on how textual autocompletion works, however, can be found in the elasticsearch documentation. Elasticsearch is an open source system that provides autocompletion capabilities out of the box. Its documentation describes how n-gram frequency statistics enable autocompletion\footnote{Elasticsearch Suggesters based on n-grams: \url{https://www.elastic.co/guide/en/elasticsearch/reference/current/search-suggesters.html}}.

\subsubsection{N-gram Frequency (non Machine Learning)}
In textual autocompletion n-gram frequency statistics are commonly applied. N-grams are substrings of a string with a length of n. For example: ``Hello World'' split in n-grams of length three, with a sliding window, would result in ``Hel'', ``ell'', ``llo'', ``lo '', and so on. Strategies might differ here, e.g. whitespaces could be removed first. Frequency statistics are obtained by counting the appearance of the n-gram across all known n-grams of search terms in the database. The frequency can be used to determine a likelihood to pick a certain search term from the database. Only search terms with a high likelihood are going to be returned as suggestions to the user. There is also already existing work on this topic \cite{mitra_query_2015}.

\subsubsection{Machine Learning}
Machine learning and neural networks, are a broad topic. Here, we will only mention which architectures can be used for generating data. Moreover, we highlight those approaches that can be utilised to extend partial text or image input.

In the field of machine learning there can be found various methods to generate text. In some cases, such approaches still need to generate n-grams first. Instead of just relying on frequency statistics, the n-grams are used to train neural networks. These neural networks are then used to generate completed versions of the partial user input. As well, there exist models that do not need to split words into n-grams at all, for instance the continuous bag of words model (CBOW) and skip-gram \cite{mikolov_efficient_2013}. Work by \citet{park_neural_2017} utilised a neural net with along short-term memory (LSTM) architecture to generate the next word of a query. The performance of a neural net depends on input data and architecture. LSTMs and recurrent neural networks (RNNs) in general, turned out to work well with text data \cite{hochreiter_long_1997}. However, latest advances in the field found Transformer networks to be superior for textbased tasks \cite{brown_language_2020, radford_language_2019, vaswani_attention_2017}. Models vary across domains. For instance, images can be automatically inpainted by utilising convolutional neural networks (CNNs) \cite{nazeri_edgeconnect_2019, wang_image_2018, yu_generative_2018}. Moreover, recent progress in image generation often uses Generative Adversarial Networks (GANs) (e.g.~\cite{karras_analyzing_2020}).

\begin{figure*}[t]
  \centering
  \includegraphics[width=0.75\textwidth]{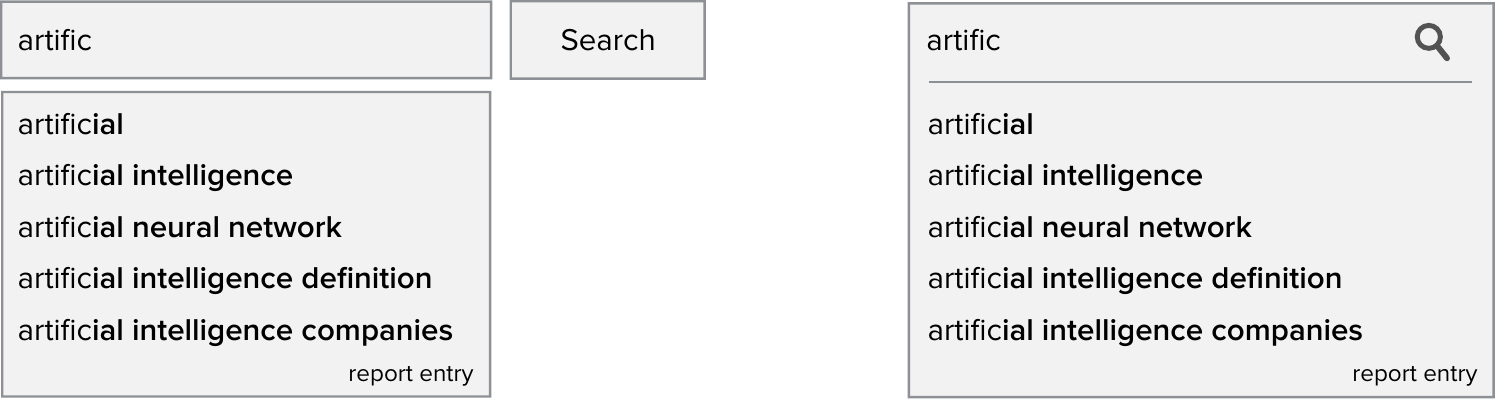}
  \caption{The figure shows the wireframes of textual autocompletion in the special case of query autocompletion. Both versions offer the same functionality. They make partial user input more complete. On the left: A user interface composed of three components, an input field, a suggestion area, and a button for confirmation. With loose interface elements. On the right: A more compact version with adapted visual composition. The search button was replaced by an icon of a magnifying glass. The search field and the suggestion area are combined into one field. Both examples show an option in the bottom right of the suggestion area to report inappropriate entries.
  }
  \label{fig:query-interface}
\end{figure*}

\section{Comparing Textual Autocompletion with Generative Approaches}

We have described autocompletion on a conceptual level and gave an introduction to technical approaches. Next, we dissect the user interface and interaction patterns of a practical example, namely textual autocompletion for search queries. This is followed by comparing and connecting textual autocompletion to generative approaches. These generative approaches are part of applications from related work or real world applications: In particular, we examine code completion\footnote{PyCharm code completion: \url{https://www.jetbrains.com/help/pycharm/auto-completing-code.html}}, sketch completion (e.g. \cite{pandian_blackbox_2020}), and layouting (e.g. \cite{dayama_grids_2020}) to textual autocompletion. 

More specifically, we first provide background information about the examples. Then we compare the user interface and finally the interaction. Parts of this comparison are similar to our prior work \cite{lehmann_autocompletion_2020}.

\subsection{Textual Autocompletion in Search Engines (Query Autocompletion)}

\textbf{User Interface:} We created a wireframe of an example of a textual autocompletion in current search engines as seen in Figure~\ref{fig:query-interface} on the left. It is a composition of three input elements. The visual design can be adapted in a way that the elements are merged and look like a single, reactive element, as seen in Figure~\ref{fig:query-interface} on the right. Yet, the basic elements remain the same. In summary, textual autocompletion consists of interface elements as follows: 

\begin{itemize}
    \item Input field for text (Input)
    \item List to display completed suggestions (Suggestion area)
    \item Button to confirm final input (Button)
\end{itemize}

Those elements can be varied across implementations. For example, the list can be displayed horizontally, instead of vertically, and it must not be a list at all. The suggestion area can be any other element as long as it is suitable to display an array of suggestions. As well, the confirm button could be hidden at the beginning and fades in after text was typed. 

\textbf{Interaction:}

\begin{figure*}[t!]
  \centering
  \includegraphics[width=1\textwidth]{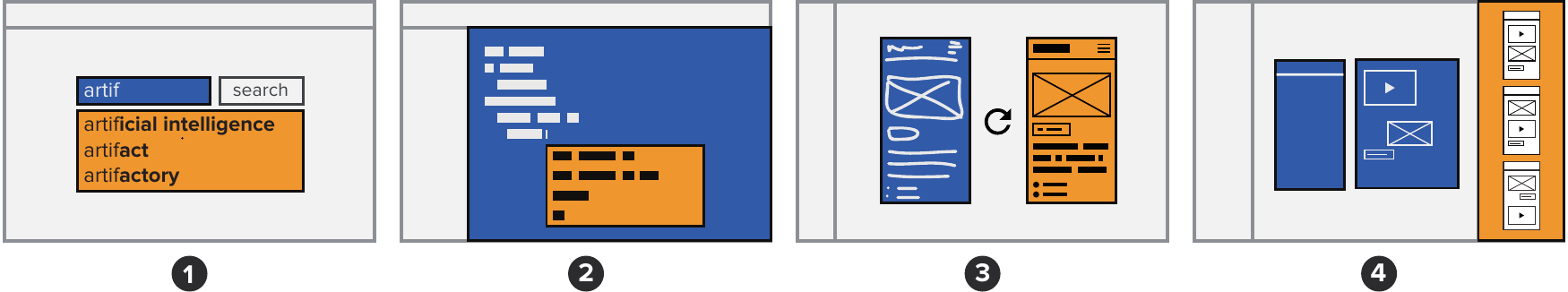}
  \caption{Comparison of the user interface of 1) search query autocompletion and three examples that share the same underlying interaction concepts, namely 2) code completion, 3) mock-up generation from sketches, and 4) layout solving. Coloured areas highlight similarities in the user interface. Blue areas with white symbols indicate fields for user input, orange areas with dark grey symbols indicate fields for completed input by the AI. Example three is inspired by \cite{pandian_blackbox_2020}, Example four is inspired by \cite{dayama_grids_2020}. All examples have the main function to make something (partial user input) more complete.
  }
  \label{fig:user-interface-comparison}
\end{figure*}

\begin{figure*}[t!]
  \centering
  \includegraphics[width=0.75\textwidth]{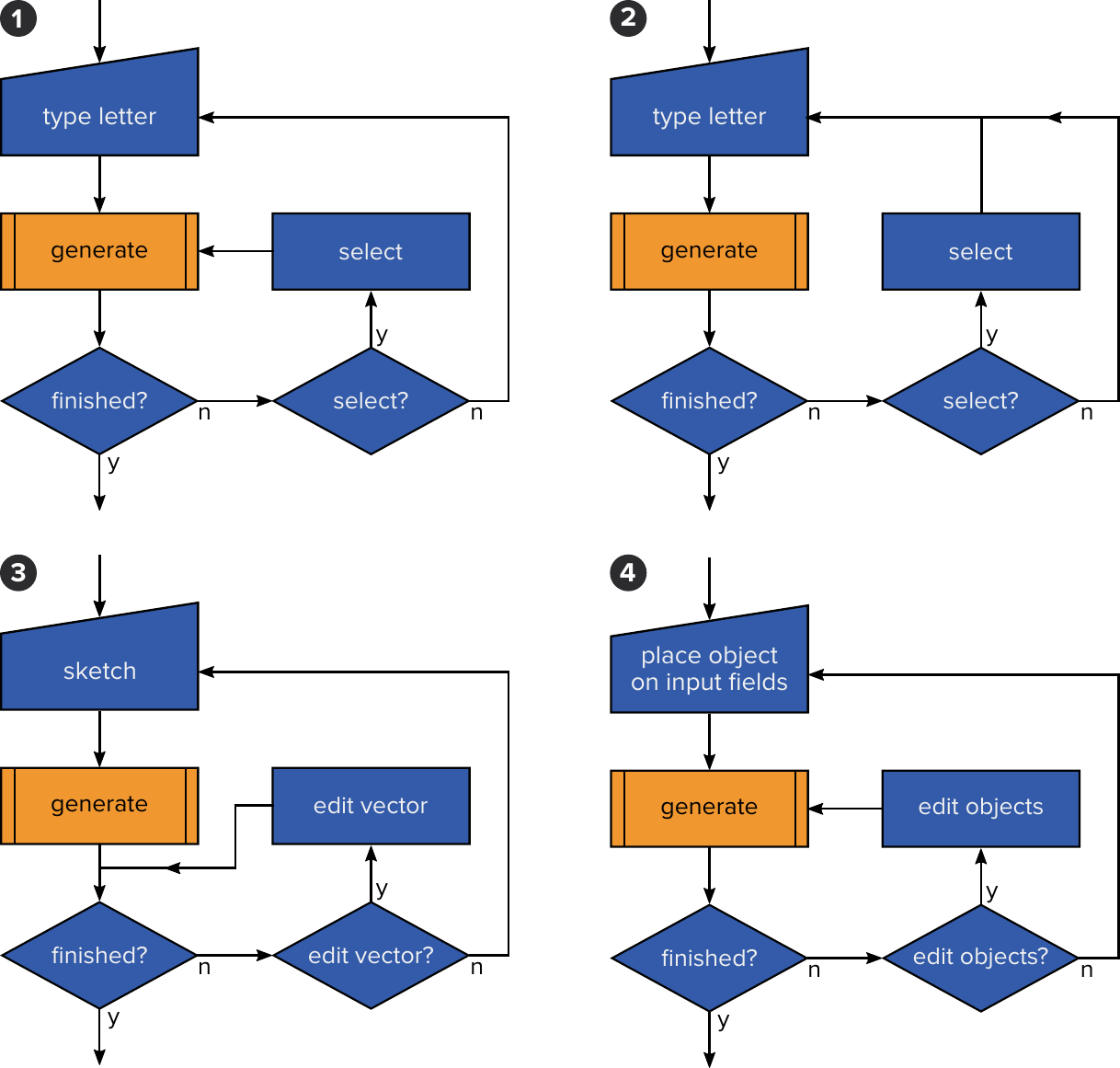}
  \caption{Visualisation of interaction patterns as flow charts. Presenting 1) search query autocompletion and generative approaches, such as 2) code completion, 3) mock-up generation from sketches, and 4) layout solving. Blue elements indicate user interaction, orange areas indicate system interaction. Chart three relates to \cite{pandian_blackbox_2020}, chart four relates to \cite{dayama_grids_2020}. Even though details in the interaction loops differ slightly, they all share the similarity to have a generative intelligent system in the loop. The generative process aims to make partial user input more complete. The final decision to accept a generated object is made by the user.}
  \label{fig:user-interaction-comparison}
\end{figure*}

The UI may be structured as seen in Figure~\ref{fig:user-interface-comparison}, example one. We observed the following interactions and put them into a flowchart as seen in Figure~\ref{fig:user-interaction-comparison}, chart one: The interaction starts with selecting the input field. Then a letter is typed with the keyboard. Next, the user can choose to confirm the input or select one of the suggestions. After selecting a suggestion, the user can again choose to confirm the input or select another suggestion. Alternatively, the text can be further edited with the keyboard by typing another letter or correcting prior input. The interaction finally ends with a confirmation. 

If a user finds a suggested entry inappropriate it can be reported through a link in the bottom right of the suggestion area. 

The interaction flow is quite simple. Yet, it offers functionality that is widely accepted for instance in search engines. Because of its simplicity we find it promising to transfer this basic flow to applications that offer human-AI interaction. As well, it holds important information: It allows for continuous interaction between the system (backend) and the user (through the frontend). Moreover, the final control is left to the user. For more complex tasks this interaction flow might have to be extended. Here, suggestions could involve all sorts of contextual data, for instance location, intent, or emotion. Output by the systems could be communicated to the user differently, for instance by an agent through conversational approaches.

\subsection{Code competion in Code Editors / IDEs}

We consider code to be a formal and structured type of text. It offers informal and structural benefits over plain text since specific parts have special functions. For example, a word could be a method, variable, or class. This is used by modern IDEs to enable for more convenient work with code. One of modern IDEs core features is code completion which is close to well-known textual autocompletion. But services like TabNine\footnote{TabNine: \url{https://www.tabnine.com}} and Kite\footnote{Kite: \url{https://www.kite.com}} offer more intelligent features, since they rely on neural nets. This is also part of current research \cite{svyatkovskiy_pythia_2019} to improve the code suggestions.

\textbf{User Interface:} A generalised user interface for autocompletion in a code editor can be seen in Figure~\ref{fig:user-interface-comparison}, example two. The user enters code into a text area (blue). Suggestions appear in a pop-up widget (orange). The widget is placed below the cursor. The widget's left edge aligns with the input cursor. The widget displays an ordered list of suggestions, e.g. method signatures. It is only visible after typing. A selected suggestion appears at the position of the input cursor.

\textbf{Interaction:} The interaction for autocompletion in code editors or IDEs is similar to textual autocompletion in search engines. The interaction flow for autocompletion is depicted in Figure~\ref{fig:user-interaction-comparison}, chart two. The interaction starts after the cursor is set in the textarea. The users enters input via the keyboard. This input is used to generate suggestions. If the input is not finished, the user can decide to select a suggestion, or ignore it. On selection, the suggested code is placed at the position of the cursor. Now, the loop starts over again.

\subsection{Intelligent UI Sketching Tools}
Contrary to code completion, image generation systems are not part of nowadays workflows in image and GUI editing tools. The latest efforts in research, however, show progress in this area. For instance, completion of partially drawn sketches \cite{tirkaz_sketched_2012} and transforming paper drawn sketches to digital wireframes \cite{buschek_paper2wire_2020}. Similar to the latter, another tool can generate medium fidelity mock-ups from low fidelity sketches \cite{pandian_blackbox_2020}. 

\textbf{User Interface:} For the description of the user interface, we orient on already existing work \cite{pandian_blackbox_2020}, and depicted the UI in Figure~\ref{fig:user-interface-comparison}, example three. The user sketches on a canvas element (blue). In fixed intervals, a medium fidelity mock-up appears on another canvas element on the right hand-side (orange). Between the two canvas elements, a button is displayed to manually trigger the generation of the medium fidelity mock-up.

\textbf{Interaction:} The user starts sketching with a digital brush or pen tool on a canvas element. That is when the interaction starts, see Figure~\ref{fig:user-interaction-comparison}, chart three. The system generates in fixed intervals a medium fidelity mock-up consisting of vector graphics. The user can then accept the mock-up by saving it. Alternatively, the user can edit the vector elements and save the mock-up afterwards. The user could also modify the sketch to modify the mock-up.

\subsection{Layout Generators}
Similar to code completion for textual autocompletion, solving layouts can be considered a specific problem within the domain of graphical user interfaces. To arrange a user interface, layout possibilities increase with the number of interface elements. There are logical constraints, however, that limit the variations. For final results, some variants are to be preferred over others. Layouting itself is a time consuming manual task. Recently, an interactive layout solving tool was introduced, on which we orient for dissecting the user interface and interaction \cite{dayama_grids_2020}.

\textbf{User Interface:} A generalised wireframe of the user interface can be seen in Figure~\ref{fig:user-interface-comparison}, example four. Pre-defined interface elements are presented in a toolbar. These elements can be dragged and placed into a workspace area (the blue area on the right-hand side). Here, the elements are placed without depending contextually on each other. On the left is another workspace area (the more narrow, blue area). However, in that area, elements are arranged to constrain the layout. A toolbar on the right (orange) displays all suggested layout solutions.

\textbf{Interaction:} The interaction starts when the user places objects on the workspaces. Compare Figure~\ref{fig:user-interaction-comparison}, chart four. After all objects have been placed, the generation is triggered by the user manually, or in fixed intervals. Subsequently, the layout solver combines the inputs and generates layout variations. The user is free to save layout suggestions and finish interaction. If the layouts are not sufficient, the user can decide to edit the objects to generate new layouts. Alternatively, the user can place new objects to start over with the interaction process.

\subsection{Comparing the UI of Generative Approaches with Query Autocompletion}
For the comparison of the user interface, we have visualised simplified wireframes, as seen in Figure~\ref{fig:user-interface-comparison}. These wireframes help us to highlight the important parts of the user interface. The blue regions indicate fields for user input. The orange fields indicate the area for system generated suggestions. We observed similarities between all interfaces. Even though layouts differ from example to example, the function remains the same. All applications provide a field for input. Here, only partial and incomplete input is done by the user. Then the system generates an extended version of it. Where possible, the system generates a set of distinct suggestions for output. The only exception is example three in Figure~\ref{fig:user-interface-comparison}, since the system outputs only one suggestion. All examples, however, share an area where the generated output is presented to the user. These areas are aligned near the input to allow the user to parse the generated output easily. Moreover, all applications use the input and output fields as the primary elements for user interaction.

\subsection{Comparing the Interaction of Generative Approaches with Query Autocompletion}
Similar to the comparison of the user interface, we have also dissected the user interactions for each example and visualised this in Figure~\ref{fig:user-interaction-comparison}. We relied on flow charts to visualise the interaction in a formal style. Blue elements indicate user interaction, orange elements indicate system operation. For our interaction flow charts, we oriented on practical examples such as Google Search, Jetbrains PyCharm IDE, and examples from related work \cite{dayama_grids_2020, pandian_blackbox_2020}. As seen in Figure~\ref{fig:user-interaction-comparison}, the interaction flows are all similar. However, they differ slightly between examples, chart one and chart two in Figure~\ref{fig:user-interaction-comparison} are good examples for these minimal differences in interaction: After selecting a suggestion in chart one, the system instantly generates another suggestion. This is different to chart two where the user needs to provide manual input before the next suggestion is generated. These are slight differences, however, they can influence the workflow crucially since both flows are very repetitive and happen very often when a user executes input.
We want to note, that our flow charts are abstract and generalised -- specific implementations use altered interaction flows. Our visualisation however, highlights that all approaches have a generative intelligent system in the loop. The user interaction starts with the operation found at the top of each flow chart. Followed by the generative process operated by the intelligent system in the backend. Afterwards, the user has to decide to accept the result. If not, the user can select a suggestion, further refine it, or ignore it and feed new partial input to the generative process. The ``finish'' switch is symbolic and indicates an end of the user interaction. This could be finalising the input, e.g. by running a search query, starting a new line in a code editor, saving a wireframe, or saving a layout. All examples are user-centred since the user is always in control and can decide whether to accept a generated object or not. In general, the role of the intelligent system is to make the user input more complete.

\begin{table*}[t]
\begin{tabularx}{\linewidth}{lX}
\toprule
Aspect         & Description      \\
\midrule
User Interface & The interface holds a field for input. Generated objects are placed near the input.                              \\
Workflow       & One or more suggestions are generated interactively and are part of the workflow                             \\
User Decision  & The user can decide to accept a suggestion or not.                                                           \\
Editing        & The user can further edit the suggestion object.                                                                \\
Information    & Partial user input serves as input for the system. The AI’s prediction is conditioned on the input to extend it.  \\
\bottomrule
\end{tabularx}
\caption{We identified five inherent aspects of autocompletion in interactive applications. Applications offering autocompletion help to make partial user input more complete. We propose that these five aspects could be transferred to other interactive AI applications that offer features to extend partial user input.}~\label{tab:autocompletion-aspects}
\end{table*}

\subsection{Aspects of Query Autocompletion}

Based on our analysis of user interfaces and interaction flows across different domains, as seen in Figure~\ref{fig:user-interface-comparison} and Figure~\ref{fig:user-interaction-comparison}, we derived five aspects that define autocompletion from a user-centred perspective. These aspects can help to inform and design novel applications which allow for human-AI interaction. 
We summarise and highlight these aspects additionally in Table~\ref{tab:autocompletion-aspects}. In the following, we describe them in more detail:

\textbf{User Interface} -- The user interface serves as an abstract visual layer that reveals the functions to the user at the frontend. This way it is rather less important for the user to understand the underlying technology in the backend. A minimalistic user interface suitable for autocompletion should hold a field for input and a suggestion area. Generated objects should be placed near, but separate from the input.

\textbf{Workflow} -- The user interfaces allow for continuous interaction between the system and the user. The system generates suggestions interactively and can become also a part of the workflow. The interaction between the system and the user is continuous until the input is finally confirmed.

\textbf{User Decision} -- The user can freely decide to accept a suggestion or not. The suggestions can be ignored which underlines the supportive and rather passive role of the system. In case that the user ignores the suggestions, the system keeps on generating new versions.

\textbf{Editing} -- The user can freely edit a taken suggestion until it fits the intent. The user can extend the suggestions, or delete them. The system should be tolerant of errors, by automatically suggesting corrections or informing the user about an error. 

\textbf{Information} -- The user input is never considered complete, always partial input is served to the system. The underlying AI (backend) is conditioned to predict a more complete version of this partial user input. Information can not only be retrieved from the explicit input, but also from other implicit variables, such as dwell time.

\section{Discussion}
Here, we reflect on the described conceptual connections, drawn between autocompletion and AI with generative capabilities, for integration of intelligent features in today's applications.

\subsection{Understanding the Concept of Autocompletion}
With our analysis, we examined autocompletion on a conceptual level. In more detail, we looked at the user interface and the interaction but gave also brief descriptions of the underlying technology. Because it is well-known from search engines, we used text query autocompletion as an example for autocompletion. Yet, our analysis goes beyond textual autocompletion by considering other domains and different approaches, such as systems that rely on machine learning methods to generate wireframes from digital sketches.

We found autocompletion to share similarities in the user interfaces across domains, as well as a basic interaction flow. We identified five aspects inherent to autocompletion applications. With our comparison of query autocompletion with other generative approaches, we demonstrated that one role of intelligent generative systems is to extend and continue partial user input.

Our research here is conceptual: We did not run a user study or analyse empirical data. We rather analysed the state-of-the-art of today's applications which integrate autocompletion. We outlined five aspects that are common to applications completing user input: The user interface, the workflow, the user decision, editing, and information. A summary can be found in Table~\ref{tab:autocompletion-aspects}. These aspects can provide a high level starting point for interaction with applications capable of generating things. We particularly expect these to prove useful for designing \textit{computational} (AI) tools since they keep the focus on designing interactions, not interfaces~\cite{beaudouin-lafon_designing_2004}.

As a key aspect of this work, we revealed analogies in graphical user interfaces and interaction flows between query (textual) autocompletion and other intelligent, generative approaches. In terms of the user interface, we looked for the function of an interface element and compared it between applications, as seen in Figure~\ref{fig:user-interface-comparison}. For this, we disregarded the visual design. For real world deployments, this means that such interface elements may look completely different but still serve the same function. Different factors could influence the visual design, such as the type of data, alignment of the interface elements, input modalities, device constraints, and so on. We suggest interpreting the user interface as a rather abstract layer that simplifies working with the underlying technology. If we can understand the function of interface elements, then we can transfer them to different domains, as recognised for autocompletion here.

For the user interaction, we explicated them using flow charts, as seen in Figure~\ref{fig:user-interaction-comparison}. The interaction between the user and intelligent systems then can be understood as a sequence of actions over time. By visually coding the operations in the flow chart, we identified an interaction loop in all examples. However, the flow charts capture only one generalised interaction flow in each example. More variants might exist, depending on the use case and implementation. Still, we suppose the user-centred interaction flow to be persistent.

In general, the detailed inspection of autocompletion here demonstrates potential benefits and insights gained from analysing and dissecting patterns in already existing interfaces, interactions, and technologies. 

\subsection{Generative Systems can be Used to Extend and Complete User Input}
We revealed an inherent role in intelligent generative systems by analysing autocompletion on a conceptual level and underlining its function in different application domains. This role is ``to extend and complete user input''. 
In the context of our work, we summarised different approaches as ``intelligent'' as long as they showed the capability to extend partial user input, and provide more completed suggestions to the user. For example, this could be statistical methods from NLP, integer programming, or machine learning. 
Similar to the examples for autocompletion we have examined in our analysis, such generative approaches from machine learning work with partial input and generate extended data. Thus, one role of generative machine learning is ``to extend and complete user input'', and it is not only limited to text data. 

Looking ahead, for example, a neural net might generate a complete text document from only a few keywords. In a scenario like this, we could assign at least one more role to generative machine learning. For instance, the role ``to inspire the user''. This would make the role of the system more active. At the same time, we would have to assume a role change in the user. The user would rather become an editor, instead of being an author. Besides text, this might also be anticipated for other domains, for example user interface design. Generative systems are able to generate functional wireframes from paper sketches \cite{buschek_paper2wire_2020}. Taking this idea to a scenario where the user provides only sketches and the system returns complete visual designs, we assume the roles would change similarly to text generation.

Systems that merely provide autocompletion, however, are less active, more passive. We consider them to be user-centred, since the user is in control of the system. The system output depends on the partial user input. Furthermore, the user is free to decide to accept a suggestion or to ignore it.

Considering the progress of machine learning over the last years it might be possible that future machine learning models will increase in performance. Their capabilities will improve, as well, tasks will be more complex. Given the techniques and computational power is evolving as in the recent past. This might open new possibilities and thus it is likely that intelligent features will be implemented in applications. We suggest to rethink the role of such systems in general, and how we want to integrate them into our workflows in the future.

\subsection{Addressing Challenges in Human-AI Interaction}

As alluded to in the intro, it is a recognised challenge to design for interactive applications of AI and recent work \cite{yang_re-examining_2020} has summarised these design challenges. Here we examine three of them to reflect on and discuss in the light of our work in this paper. 

\subsubsection{The Challenge of Envisioning Interaction with AI} Similar to our investigation of autocompletion in this paper, we suggest to analyse already existing intelligent tools for inspiration. Ideally, these can be dissected into reusable interface and interaction patterns, as shown in our example here. Developing such a set of interface and interaction patterns over time might then facilitate composing new interactions with intelligent systems.

\subsubsection{The Challenge of Understanding AI Capabilities} The autocompletion pattern might also facilitate understanding of AI, concretely, by putting input-output mappings at the core of the interaction, thus making them \textit{explorable}. First, the AI is fed with partial user input, which the user can quickly vary and iterate on to explore ``AI reactions'' and thus potentially develop a (tacit) understanding of it, possibly similar to experiences with rule-based systems. Second, autocompletion typically generates \textit{multiple} variants as output. This might help the user to judge AI capabilities, since it gives a glimpse at the AI's potential output space, especially also across repeated input variation/iteration. Third, autocompletion typically \textit{ranks} output, for example by probability, which might facilitate user understanding of AI capabilities as it gives a simple way of directly indicating the AI's (relative) uncertainty in the GUI.

\subsubsection{The Challenge of Crafting Interactions for Unpredictable Output} The output of intelligent systems can be unpredictable from time to time. For instance, intelligent systems might generate text or images that are inappropriate to the user. Autocompletion provides an example UI for addressing this challenge in three ways: First, by design, it leaves the final decision about the accepted content to the user. Second, it shows multiple options, thus possibly including not only an ``outlier'' but also more appropriate alternatives (or at least supporting the discovery of an ``outlier'' as such). Third, as depicted in Figure~\ref{fig:query-interface}, the autocomplete UI easily affords to give explicit feedback on inappropriate output. This way, a personalised filter could be created over time or the signals could be used for the next training iteration of a machine learning model to keep out inappropriate output in the future. 

In summary, our suggestions on the design challenges illustrate how a well-known UI/interaction concept such as autocompletion can be used as a conceptual lens and starting points to design for interactive intelligent systems.

For realising this potential in practice we see the key in collaborations between domain experts: This could help to integrate intelligent features in future workflows within applications. Moreover, this might simplify practical work with machine learning as a design material. This is important since machine learning is supposed to add complexity to software architecture and interfaces at the same time (e.g.~\citet{yang_planning_2016}). This trade-off between functional complexity and a simplistic user interface should be addressed when designing for intelligent systems. The complexity should be reduced at least for the user of a human-AI application.

\subsection{Mixed-Initiative User Interfaces for Human-AI Interaction}
Beyond these design challenges, intelligent applications offer great potential, for example, to support finding ideas \cite{alvarez_fostering_2018, liapis_can_2016, yannakakis_mixed-initiative_2014}. In general, we see the opportunity to connect the ideas of this work. For instance, to sense of agency. Especially for user-centred approaches, it is important to measure how much the user feels in control over a tool. Another aspect that deserves more attention is timing of AI capabilities in interactive use. For example, timing of updates in autocompletion are driven by the user (e.g. typing another character triggers updated completions). However, one might study when to offer autocompletion at all (e.g. for text completion), since it also requires attention (cf.~\cite{quinn_cost-benefit_2016}). This could be combined with research on negative feedback and error-tolerance. Negative feedback could be used to infer actions to adapt the AI involvement at the interface level accordingly.

Both timing and sense of agency could be examined in particular in light of the concept of mixed initiative interfaces~\cite{horvitz_principles_1999}. Our analysis of autocompletion also already connects to this -- both user and AI system contribute to the emerging digital content (e.g. query, text, image) via a specific input-generation-selection loop (Figure~\ref{fig:user-interaction-comparison}). 

\section{Conclusion}
With this paper, we examined autocompletion on a conceptual level and analysed its interface, interaction, and technical elements. We identified reoccurring interface and interaction patterns in autocompletion across several domains, in particular going beyond the ``traditional'' text query completion. For example, we recognised analogies to autocompletion in AI-support for digital sketches and layouting. Based on our conceptual analysis, we suggested and discussed autocompletion as an inspiration and conceptual lens on current challenges in designing for human-AI interaction. With  this work, we hope to provide a pragmatic, concrete conceptual starting point to help envision interaction designs with and for AI that can generate new things. 

As future work, we plan to conduct experimental studies to empirically investigate the transfer and use of autocomplete UIs for interaction with generative AI as conceptually extracted here. More broadly, the highlighted inherent aspects of the autocomplete pattern further motivate investigations in combination with topics from mixed initiative interaction, sense of agency, and timing.

\begin{acks}
  This project is funded by the Bavarian State Ministry of Science and the Arts and coordinated by the Bavarian Research Institute for Digital Transformation (bidt).
\end{acks}

\bibliographystyle{ACM-Reference-Format}
\bibliography{sample-base}


\begin{thebibliography}{50}


\ifx \showCODEN    \undefined \def \showCODEN     #1{\unskip}     \fi
\ifx \showDOI      \undefined \def \showDOI       #1{#1}\fi
\ifx \showISBNx    \undefined \def \showISBNx     #1{\unskip}     \fi
\ifx \showISBNxiii \undefined \def \showISBNxiii  #1{\unskip}     \fi
\ifx \showISSN     \undefined \def \showISSN      #1{\unskip}     \fi
\ifx \showLCCN     \undefined \def \showLCCN      #1{\unskip}     \fi
\ifx \shownote     \undefined \def \shownote      #1{#1}          \fi
\ifx \showarticletitle \undefined \def \showarticletitle #1{#1}   \fi
\ifx \showURL      \undefined \def \showURL       {\relax}        \fi
\providecommand\bibfield[2]{#2}
\providecommand\bibinfo[2]{#2}
\providecommand\natexlab[1]{#1}
\providecommand\showeprint[2][]{arXiv:#2}

\bibitem[Alvarez et~al\mbox{.}(2018)]%
        {alvarez_fostering_2018}
\bibfield{author}{\bibinfo{person}{Alberto Alvarez}, \bibinfo{person}{Steve
  Dahlskog}, \bibinfo{person}{Jose Font}, \bibinfo{person}{Johan Holmberg},
  \bibinfo{person}{Chelsi Nolasco}, {and} \bibinfo{person}{Axel Österman}.}
  \bibinfo{year}{2018}\natexlab{}.
\newblock \showarticletitle{Fostering creativity in the mixed-initiative
  evolutionary dungeon designer}. In \bibinfo{booktitle}{\emph{Proceedings of
  the 13th {International} {Conference} on the {Foundations} of {Digital}
  {Games}}}. \bibinfo{publisher}{ACM}, \bibinfo{address}{Malmö Sweden},
  \bibinfo{pages}{1--8}.
\newblock
\showISBNx{978-1-4503-6571-0}
\urldef\tempurl%
\url{https://doi.org/10.1145/3235765.3235815}
\showDOI{\tempurl}


\bibitem[Amin et~al\mbox{.}(2009)]%
        {boughanem_organizing_2009}
\bibfield{author}{\bibinfo{person}{Alia Amin}, \bibinfo{person}{Michiel
  Hildebrand}, \bibinfo{person}{Jacco van Ossenbruggen},
  \bibinfo{person}{Vanessa Evers}, {and} \bibinfo{person}{Lynda Hardman}.}
  \bibinfo{year}{2009}\natexlab{}.
\newblock \showarticletitle{Organizing {Suggestions} in {Autocompletion}
  {Interfaces}}.
\newblock In \bibinfo{booktitle}{\emph{Advances in {Information} {Retrieval}}},
  \bibfield{editor}{\bibinfo{person}{Mohand Boughanem},
  \bibinfo{person}{Catherine Berrut}, \bibinfo{person}{Josiane Mothe}, {and}
  \bibinfo{person}{Chantal Soule-Dupuy}} (Eds.). Vol.~\bibinfo{volume}{5478}.
  \bibinfo{publisher}{Springer Berlin Heidelberg}, \bibinfo{address}{Berlin,
  Heidelberg}, \bibinfo{pages}{521--529}.
\newblock
\showISBNx{978-3-642-00957-0 978-3-642-00958-7}
\urldef\tempurl%
\url{https://doi.org/10.1007/978-3-642-00958-7_46}
\showDOI{\tempurl}
\newblock
\shownote{Series Title: Lecture Notes in Computer Science}.


\bibitem[Banovic et~al\mbox{.}(2019)]%
        {banovic_limits_2019}
\bibfield{author}{\bibinfo{person}{Nikola Banovic}, \bibinfo{person}{Ticha
  Sethapakdi}, \bibinfo{person}{Yasasvi Hari}, \bibinfo{person}{Anind~K. Dey},
  {and} \bibinfo{person}{Jennifer Mankoff}.} \bibinfo{year}{2019}\natexlab{}.
\newblock \showarticletitle{The {Limits} of {Expert} {Text} {Entry} {Speed} on
  {Mobile} {Keyboards} with {Autocorrect}}. In
  \bibinfo{booktitle}{\emph{Proceedings of the 21st {International}
  {Conference} on {Human}-{Computer} {Interaction} with {Mobile} {Devices} and
  {Services}}}. \bibinfo{publisher}{ACM}, \bibinfo{address}{Taipei Taiwan},
  \bibinfo{pages}{1--12}.
\newblock
\showISBNx{978-1-4503-6825-4}
\urldef\tempurl%
\url{https://doi.org/10.1145/3338286.3340126}
\showDOI{\tempurl}


\bibitem[Bar-Yossef and Kraus(2011)]%
        {bar-yossef_context-sensitive_2011}
\bibfield{author}{\bibinfo{person}{Ziv Bar-Yossef} {and} \bibinfo{person}{Naama
  Kraus}.} \bibinfo{year}{2011}\natexlab{}.
\newblock \showarticletitle{Context-sensitive query auto-completion}. In
  \bibinfo{booktitle}{\emph{Proceedings of the 20th international conference on
  {World} wide web}} \emph{(\bibinfo{series}{{WWW} '11})}.
  \bibinfo{publisher}{ACM Press}, \bibinfo{address}{New York, NY, USA},
  \bibinfo{pages}{107--116}.
\newblock
\showISBNx{978-1-4503-0632-4}
\urldef\tempurl%
\url{https://doi.org/10.1145/1963405.1963424}
\showDOI{\tempurl}


\bibitem[Bast and Weber(2006)]%
        {bast_type_2006}
\bibfield{author}{\bibinfo{person}{Holger Bast} {and} \bibinfo{person}{Ingmar
  Weber}.} \bibinfo{year}{2006}\natexlab{}.
\newblock \showarticletitle{Type less, find more: fast autocompletion search
  with a succinct index}. In \bibinfo{booktitle}{\emph{Proceedings of the 29th
  annual international {ACM} {SIGIR} conference on {Research} and development
  in information retrieval}} \emph{(\bibinfo{series}{{SIGIR} '06})}.
  \bibinfo{publisher}{ACM Press}, \bibinfo{address}{New York, NY, USA},
  \bibinfo{pages}{364--371}.
\newblock
\showISBNx{978-1-59593-369-0}
\urldef\tempurl%
\url{https://doi.org/10.1145/1148170.1148234}
\showDOI{\tempurl}


\bibitem[Beaudouin-Lafon(2004)]%
        {beaudouin-lafon_designing_2004}
\bibfield{author}{\bibinfo{person}{Michel Beaudouin-Lafon}.}
  \bibinfo{year}{2004}\natexlab{}.
\newblock \showarticletitle{Designing interaction, not interfaces}. In
  \bibinfo{booktitle}{\emph{Proceedings of the working conference on {Advanced}
  visual interfaces - {AVI} '04}}. \bibinfo{publisher}{ACM Press},
  \bibinfo{address}{New York, NY, USA}, \bibinfo{pages}{15}.
\newblock
\showISBNx{978-1-58113-867-2}
\urldef\tempurl%
\url{https://doi.org/10.1145/989863.989865}
\showDOI{\tempurl}


\bibitem[Bennett et~al\mbox{.}(2011)]%
        {bennett_simpleflow_2011}
\bibfield{author}{\bibinfo{person}{Mike Bennett}, \bibinfo{person}{Kevin
  McCarthy}, \bibinfo{person}{Sile O’Modhrain}, {and} \bibinfo{person}{Barry
  Smyth}.} \bibinfo{year}{2011}\natexlab{}.
\newblock \showarticletitle{{SimpleFlow}: {Enhancing} {Gestural} {Interaction}
  with {Gesture} {Prediction}, {Abbreviation} and {Autocompletion}}. In
  \bibinfo{booktitle}{\emph{Human-{Computer} {Interaction} – {INTERACT}
  2011}} \emph{(\bibinfo{series}{Lecture {Notes} in {Computer} {Science}})},
  \bibfield{editor}{\bibinfo{person}{Pedro Campos}, \bibinfo{person}{Nicholas
  Graham}, \bibinfo{person}{Joaquim Jorge}, \bibinfo{person}{Nuno Nunes},
  \bibinfo{person}{Philippe Palanque}, {and} \bibinfo{person}{Marco Winckler}}
  (Eds.). \bibinfo{publisher}{Springer}, \bibinfo{address}{Berlin, Heidelberg},
  \bibinfo{pages}{591--608}.
\newblock
\showISBNx{978-3-642-23774-4}
\urldef\tempurl%
\url{https://doi.org/10.1007/978-3-642-23774-4_47}
\showDOI{\tempurl}


\bibitem[Brown et~al\mbox{.}(2020)]%
        {brown_language_2020}
\bibfield{author}{\bibinfo{person}{Tom~B. Brown}, \bibinfo{person}{Benjamin
  Mann}, \bibinfo{person}{Nick Ryder}, \bibinfo{person}{Melanie Subbiah},
  \bibinfo{person}{Jared Kaplan}, \bibinfo{person}{Prafulla Dhariwal},
  \bibinfo{person}{Arvind Neelakantan}, \bibinfo{person}{Pranav Shyam},
  \bibinfo{person}{Girish Sastry}, \bibinfo{person}{Amanda Askell},
  \bibinfo{person}{Sandhini Agarwal}, \bibinfo{person}{Ariel Herbert-Voss},
  \bibinfo{person}{Gretchen Krueger}, \bibinfo{person}{Tom Henighan},
  \bibinfo{person}{Rewon Child}, \bibinfo{person}{Aditya Ramesh},
  \bibinfo{person}{Daniel~M. Ziegler}, \bibinfo{person}{Jeffrey Wu},
  \bibinfo{person}{Clemens Winter}, \bibinfo{person}{Christopher Hesse},
  \bibinfo{person}{Mark Chen}, \bibinfo{person}{Eric Sigler},
  \bibinfo{person}{Mateusz Litwin}, \bibinfo{person}{Scott Gray},
  \bibinfo{person}{Benjamin Chess}, \bibinfo{person}{Jack Clark},
  \bibinfo{person}{Christopher Berner}, \bibinfo{person}{Sam McCandlish},
  \bibinfo{person}{Alec Radford}, \bibinfo{person}{Ilya Sutskever}, {and}
  \bibinfo{person}{Dario Amodei}.} \bibinfo{year}{2020}\natexlab{}.
\newblock \showarticletitle{Language {Models} are {Few}-{Shot} {Learners}}.
\newblock \bibinfo{journal}{\emph{arXiv:2005.14165 [cs]}} (\bibinfo{date}{July}
  \bibinfo{year}{2020}).
\newblock
\urldef\tempurl%
\url{http://arxiv.org/abs/2005.14165}
\showURL{%
\tempurl}
\newblock
\shownote{arXiv: 2005.14165}.


\bibitem[Buschek et~al\mbox{.}(2020)]%
        {buschek_paper2wire_2020}
\bibfield{author}{\bibinfo{person}{Daniel Buschek}, \bibinfo{person}{Charlotte
  Anlauff}, {and} \bibinfo{person}{Florian Lachner}.}
  \bibinfo{year}{2020}\natexlab{}.
\newblock \showarticletitle{{Paper2Wire}: a case study of user-centred
  development of machine learning tools for {UX} designers}. In
  \bibinfo{booktitle}{\emph{Proceedings of the {Conference} on {Mensch} und
  {Computer}}} \emph{(\bibinfo{series}{{MuC} '20})}.
  \bibinfo{publisher}{Association for Computing Machinery},
  \bibinfo{address}{New York, NY, USA}, \bibinfo{pages}{33--41}.
\newblock
\showISBNx{978-1-4503-7540-5}
\urldef\tempurl%
\url{https://doi.org/10.1145/3404983.3405506}
\showDOI{\tempurl}


\bibitem[Cai and de~Rijke(2016)]%
        {cai_selectively_2016}
\bibfield{author}{\bibinfo{person}{Fei Cai} {and} \bibinfo{person}{Maarten de
  Rijke}.} \bibinfo{year}{2016}\natexlab{}.
\newblock \showarticletitle{Selectively {Personalizing} {Query}
  {Auto}-{Completion}}. In \bibinfo{booktitle}{\emph{Proceedings of the 39th
  {International} {ACM} {SIGIR} conference on {Research} and {Development} in
  {Information} {Retrieval}}} \emph{(\bibinfo{series}{{SIGIR} '16})}.
  \bibinfo{publisher}{ACM Press}, \bibinfo{address}{New York, NY, USA},
  \bibinfo{pages}{993--996}.
\newblock
\showISBNx{978-1-4503-4069-4}
\urldef\tempurl%
\url{https://doi.org/10.1145/2911451.2914686}
\showDOI{\tempurl}


\bibitem[Cai et~al\mbox{.}(2016)]%
        {cai_diversifying_2016}
\bibfield{author}{\bibinfo{person}{Fei Cai}, \bibinfo{person}{Ridho Reinanda},
  {and} \bibinfo{person}{Maarten~De Rijke}.} \bibinfo{year}{2016}\natexlab{}.
\newblock \showarticletitle{Diversifying {Query} {Auto}-{Completion}}.
\newblock \bibinfo{journal}{\emph{ACM Transactions on Information Systems}}
  \bibinfo{volume}{34}, \bibinfo{number}{4} (\bibinfo{date}{June}
  \bibinfo{year}{2016}), \bibinfo{pages}{25:1--25:33}.
\newblock
\showISSN{1046-8188}
\urldef\tempurl%
\url{https://doi.org/10.1145/2910579}
\showDOI{\tempurl}


\bibitem[Cai and Rijke(2016)]%
        {cai_survey_2016}
\bibfield{author}{\bibinfo{person}{Fei Cai} {and} \bibinfo{person}{Maarten~de
  Rijke}.} \bibinfo{year}{2016}\natexlab{}.
\newblock \showarticletitle{A {Survey} of {Query} {Auto} {Completion} in
  {Information} {Retrieval}}.
\newblock \bibinfo{journal}{\emph{Foundations and Trends® in Information
  Retrieval}} \bibinfo{volume}{10}, \bibinfo{number}{4} (\bibinfo{date}{Sept.}
  \bibinfo{year}{2016}), \bibinfo{pages}{273--363}.
\newblock
\showISSN{1554-0669, 1554-0677}
\urldef\tempurl%
\url{https://doi.org/10.1561/1500000055}
\showDOI{\tempurl}
\newblock
\shownote{Publisher: Now Publishers, Inc.}.


\bibitem[Costagliola et~al\mbox{.}(2013)]%
        {costagliola_investigating_2013}
\bibfield{author}{\bibinfo{person}{Gennaro Costagliola},
  \bibinfo{person}{Mattia~De Rosa}, {and} \bibinfo{person}{Vittorio Fuccella}.}
  \bibinfo{year}{2013}\natexlab{}.
\newblock \showarticletitle{Investigating {Human} {Performance} in
  {Hand}-{Drawn} {Symbol} {Autocompletion}}. In \bibinfo{booktitle}{\emph{2013
  {IEEE} {International} {Conference} on {Systems}, {Man}, and {Cybernetics}}}.
  \bibinfo{pages}{279--284}.
\newblock
\urldef\tempurl%
\url{https://doi.org/10.1109/SMC.2013.54}
\showDOI{\tempurl}
\newblock
\shownote{ISSN: 1062-922X}.


\bibitem[Dayama et~al\mbox{.}(2020)]%
        {dayama_grids_2020}
\bibfield{author}{\bibinfo{person}{Niraj~Ramesh Dayama},
  \bibinfo{person}{Kashyap Todi}, \bibinfo{person}{Taru Saarelainen}, {and}
  \bibinfo{person}{Antti Oulasvirta}.} \bibinfo{year}{2020}\natexlab{}.
\newblock \showarticletitle{{GRIDS}: {Interactive} {Layout} {Design} with
  {Integer} {Programming}}. In \bibinfo{booktitle}{\emph{Proceedings of the
  2020 {CHI} {Conference} on {Human} {Factors} in {Computing} {Systems}}}
  \emph{(\bibinfo{series}{{CHI} '20})}. \bibinfo{publisher}{ACM Press},
  \bibinfo{address}{New York, NY, USA}, \bibinfo{pages}{1--13}.
\newblock
\showISBNx{978-1-4503-6708-0}
\urldef\tempurl%
\url{https://doi.org/10.1145/3313831.3376553}
\showDOI{\tempurl}


\bibitem[Di~Santo et~al\mbox{.}(2015)]%
        {di_santo_comparing_2015}
\bibfield{author}{\bibinfo{person}{Giovanni Di~Santo}, \bibinfo{person}{Richard
  McCreadie}, \bibinfo{person}{Craig Macdonald}, {and} \bibinfo{person}{Iadh
  Ounis}.} \bibinfo{year}{2015}\natexlab{}.
\newblock \showarticletitle{Comparing {Approaches} for {Query}
  {Autocompletion}}. In \bibinfo{booktitle}{\emph{Proceedings of the 38th
  {International} {ACM} {SIGIR} {Conference} on {Research} and {Development} in
  {Information} {Retrieval}}} \emph{(\bibinfo{series}{{SIGIR} '15})}.
  \bibinfo{publisher}{ACM Press}, \bibinfo{address}{New York, NY, USA},
  \bibinfo{pages}{775--778}.
\newblock
\showISBNx{978-1-4503-3621-5}
\urldef\tempurl%
\url{https://doi.org/10.1145/2766462.2767829}
\showDOI{\tempurl}


\bibitem[Hochreiter and Schmidhuber(1997)]%
        {hochreiter_long_1997}
\bibfield{author}{\bibinfo{person}{Sepp Hochreiter} {and}
  \bibinfo{person}{Jürgen Schmidhuber}.} \bibinfo{year}{1997}\natexlab{}.
\newblock \showarticletitle{Long {Short}-{Term} {Memory}}.
\newblock \bibinfo{journal}{\emph{Neural Computation}} \bibinfo{volume}{9},
  \bibinfo{number}{8} (\bibinfo{date}{Nov.} \bibinfo{year}{1997}),
  \bibinfo{pages}{1735--1780}.
\newblock
\showISSN{0899-7667, 1530-888X}
\urldef\tempurl%
\url{https://doi.org/10.1162/neco.1997.9.8.1735}
\showDOI{\tempurl}


\bibitem[Hofmann et~al\mbox{.}(2014)]%
        {hofmann_eye-tracking_2014}
\bibfield{author}{\bibinfo{person}{Kajta Hofmann}, \bibinfo{person}{Bhaskar
  Mitra}, \bibinfo{person}{Filip Radlinski}, {and} \bibinfo{person}{Milad
  Shokouhi}.} \bibinfo{year}{2014}\natexlab{}.
\newblock \showarticletitle{An {Eye}-tracking {Study} of {User} {Interactions}
  with {Query} {Auto} {Completion}}. In \bibinfo{booktitle}{\emph{Proceedings
  of the 23rd {ACM} {International} {Conference} on {Conference} on
  {Information} and {Knowledge} {Management}}} \emph{(\bibinfo{series}{{CIKM}
  '14})}. \bibinfo{publisher}{ACM Press}, \bibinfo{address}{New York, NY, USA},
  \bibinfo{pages}{549--558}.
\newblock
\showISBNx{978-1-4503-2598-1}
\urldef\tempurl%
\url{https://doi.org/10.1145/2661829.2661922}
\showDOI{\tempurl}


\bibitem[Horvitz(1999)]%
        {horvitz_principles_1999}
\bibfield{author}{\bibinfo{person}{Eric Horvitz}.}
  \bibinfo{year}{1999}\natexlab{}.
\newblock \showarticletitle{Principles of mixed-initiative user interfaces}. In
  \bibinfo{booktitle}{\emph{Proceedings of the {SIGCHI} conference on {Human}
  {Factors} in {Computing} {Systems}}} \emph{(\bibinfo{series}{{CHI} '99})}.
  \bibinfo{publisher}{ACM Press}, \bibinfo{address}{New York, NY, USA},
  \bibinfo{pages}{159--166}.
\newblock
\showISBNx{978-0-201-48559-2}
\urldef\tempurl%
\url{https://doi.org/10.1145/302979.303030}
\showDOI{\tempurl}


\bibitem[Hsu et~al\mbox{.}(2020)]%
        {hsu_autocomplete_2020}
\bibfield{author}{\bibinfo{person}{Chen-Yuan Hsu}, \bibinfo{person}{Li-Yi Wei},
  \bibinfo{person}{Lihua You}, {and} \bibinfo{person}{Jian~Jun Zhang}.}
  \bibinfo{year}{2020}\natexlab{}.
\newblock \showarticletitle{Autocomplete {Element} {Fields}}. In
  \bibinfo{booktitle}{\emph{Proceedings of the 2020 {CHI} {Conference} on
  {Human} {Factors} in {Computing} {Systems}}}. \bibinfo{publisher}{ACM Press},
  \bibinfo{address}{New York, NY, USA}, \bibinfo{pages}{1--13}.
\newblock
\showISBNx{978-1-4503-6708-0}
\urldef\tempurl%
\url{https://doi.org/10.1145/3313831.3376248}
\showDOI{\tempurl}


\bibitem[Huang and Canny(2019)]%
        {huang_sketchforme_2019}
\bibfield{author}{\bibinfo{person}{Forrest Huang} {and}
  \bibinfo{person}{John~F. Canny}.} \bibinfo{year}{2019}\natexlab{}.
\newblock \showarticletitle{Sketchforme: {Composing} {Sketched} {Scenes} from
  {Text} {Descriptions} for {Interactive} {Applications}}. In
  \bibinfo{booktitle}{\emph{Proceedings of the 32nd {Annual} {ACM} {Symposium}
  on {User} {Interface} {Software} and {Technology}}}.
  \bibinfo{publisher}{ACM}, \bibinfo{address}{New York, NY, USA},
  \bibinfo{pages}{209--220}.
\newblock
\showISBNx{978-1-4503-6816-2}
\urldef\tempurl%
\url{https://doi.org/10.1145/3332165.3347878}
\showDOI{\tempurl}


\bibitem[Jakobsson(1986)]%
        {jakobsson_autocompletion_1986}
\bibfield{author}{\bibinfo{person}{M. Jakobsson}.}
  \bibinfo{year}{1986}\natexlab{}.
\newblock \showarticletitle{Autocompletion in full text transaction entry: a
  method for humanized input}.
\newblock \bibinfo{journal}{\emph{ACM SIGCHI Bulletin}} \bibinfo{volume}{17},
  \bibinfo{number}{4} (\bibinfo{date}{April} \bibinfo{year}{1986}),
  \bibinfo{pages}{327--332}.
\newblock
\showISSN{0736-6906}
\urldef\tempurl%
\url{https://doi.org/10.1145/22339.22391}
\showDOI{\tempurl}


\bibitem[Karras et~al\mbox{.}(2020)]%
        {karras_analyzing_2020}
\bibfield{author}{\bibinfo{person}{Tero Karras}, \bibinfo{person}{Samuli
  Laine}, \bibinfo{person}{Miika Aittala}, \bibinfo{person}{Janne Hellsten},
  \bibinfo{person}{Jaakko Lehtinen}, {and} \bibinfo{person}{Timo Aila}.}
  \bibinfo{year}{2020}\natexlab{}.
\newblock \showarticletitle{Analyzing and {Improving} the {Image} {Quality} of
  {StyleGAN}}. In \bibinfo{booktitle}{\emph{Proceedings of the {IEEE}/{CVF}
  {Conference} on {Computer} {Vision} and {Pattern} {Recognition} ({CVPR})}}.
  \bibinfo{publisher}{IEEE}, \bibinfo{address}{Seattle, WA, USA},
  \bibinfo{pages}{8107--8116}.
\newblock
\showISBNx{978-1-72817-168-5}
\urldef\tempurl%
\url{https://doi.org/10.1109/CVPR42600.2020.00813}
\showDOI{\tempurl}


\bibitem[Lehmann and Buschek(2020)]%
        {lehmann_autocompletion_2020}
\bibfield{author}{\bibinfo{person}{Florian Lehmann} {and}
  \bibinfo{person}{Daniel Buschek}.} \bibinfo{year}{2020}\natexlab{}.
\newblock \showarticletitle{Autocompletion as a {Basic} {Interaction} {Concept}
  for {User}-{Centered} {AI}}.
\newblock  (\bibinfo{year}{2020}).
\newblock
\urldef\tempurl%
\url{https://doi.org/10.18420/MUC2020-WS111-328}
\showDOI{\tempurl}
\newblock
\shownote{Publisher: Gesellschaft für Informatik e.V.}.


\bibitem[Li et~al\mbox{.}(2015)]%
        {li_analyzing_2015}
\bibfield{author}{\bibinfo{person}{Liangda Li}, \bibinfo{person}{Hongbo Deng},
  \bibinfo{person}{Anlei Dong}, \bibinfo{person}{Yi Chang},
  \bibinfo{person}{Hongyuan Zha}, {and} \bibinfo{person}{Ricardo Baeza-Yates}.}
  \bibinfo{year}{2015}\natexlab{}.
\newblock \showarticletitle{Analyzing {User}'s {Sequential} {Behavior} in
  {Query} {Auto}-{Completion} via {Markov} {Processes}}. In
  \bibinfo{booktitle}{\emph{Proceedings of the 38th {International} {ACM}
  {SIGIR} {Conference} on {Research} and {Development} in {Information}
  {Retrieval}}} \emph{(\bibinfo{series}{{SIGIR} '15})}. \bibinfo{publisher}{ACM
  Press}, \bibinfo{address}{New York, NY, USA}, \bibinfo{pages}{123--132}.
\newblock
\showISBNx{978-1-4503-3621-5}
\urldef\tempurl%
\url{https://doi.org/10.1145/2766462.2767723}
\showDOI{\tempurl}


\bibitem[Li et~al\mbox{.}(2014)]%
        {li_two-dimensional_2014}
\bibfield{author}{\bibinfo{person}{Yanen Li}, \bibinfo{person}{Anlei Dong},
  \bibinfo{person}{Hongning Wang}, \bibinfo{person}{Hongbo Deng},
  \bibinfo{person}{Yi Chang}, {and} \bibinfo{person}{ChengXiang Zhai}.}
  \bibinfo{year}{2014}\natexlab{}.
\newblock \showarticletitle{A two-dimensional click model for query
  auto-completion}. In \bibinfo{booktitle}{\emph{Proceedings of the 37th
  international {ACM} {SIGIR} conference on {Research} \& development in
  information retrieval}} \emph{(\bibinfo{series}{{SIGIR} '14})}.
  \bibinfo{publisher}{ACM Press}, \bibinfo{address}{New York, NY, USA},
  \bibinfo{pages}{455--464}.
\newblock
\showISBNx{978-1-4503-2257-7}
\urldef\tempurl%
\url{https://doi.org/10.1145/2600428.2609571}
\showDOI{\tempurl}


\bibitem[Liapis(2016)]%
        {liapis_can_2016}
\bibfield{author}{\bibinfo{person}{Antonios Liapis}.}
  \bibinfo{year}{2016}\natexlab{}.
\newblock \showarticletitle{Can computers foster human users' creativity?
  theory and praxis of mixed-initiative co-creativity}.
\newblock \bibinfo{journal}{\emph{Digital Culture \& Education}}
  \bibinfo{volume}{8(2)} (\bibinfo{year}{2016}), \bibinfo{pages}{136--153}.
\newblock
\urldef\tempurl%
\url{https://www.um.edu.mt/library/oar/handle/123456789/29476}
\showURL{%
\tempurl}


\bibitem[Lin et~al\mbox{.}(2012)]%
        {lin_lotusx_2012}
\bibfield{author}{\bibinfo{person}{Chunbin Lin}, \bibinfo{person}{Jiaheng Lu},
  \bibinfo{person}{Tok~Wang Ling}, {and} \bibinfo{person}{Bogdan Cautis}.}
  \bibinfo{year}{2012}\natexlab{}.
\newblock \showarticletitle{{LotusX}: {A} {Position}-{Aware} {XML} {Graphical}
  {Search} {System} with {Auto}-{Completion}}. In
  \bibinfo{booktitle}{\emph{2012 {IEEE} 28th {International} {Conference} on
  {Data} {Engineering}}}. \bibinfo{publisher}{IEEE},
  \bibinfo{address}{Washington, DC, USA}, \bibinfo{pages}{1265--1268}.
\newblock
\urldef\tempurl%
\url{https://doi.org/10.1109/ICDE.2012.123}
\showDOI{\tempurl}
\newblock
\shownote{ISSN: 2375-026X}.


\bibitem[Mikolov et~al\mbox{.}(2013)]%
        {mikolov_efficient_2013}
\bibfield{author}{\bibinfo{person}{Tomas Mikolov}, \bibinfo{person}{Kai Chen},
  \bibinfo{person}{Greg Corrado}, {and} \bibinfo{person}{Jeffrey Dean}.}
  \bibinfo{year}{2013}\natexlab{}.
\newblock \showarticletitle{Efficient {Estimation} of {Word} {Representations}
  in {Vector} {Space}}.
\newblock \bibinfo{journal}{\emph{arXiv:1301.3781 [cs]}} (\bibinfo{date}{Sept.}
  \bibinfo{year}{2013}).
\newblock
\urldef\tempurl%
\url{http://arxiv.org/abs/1301.3781}
\showURL{%
\tempurl}
\newblock
\shownote{arXiv: 1301.3781}.


\bibitem[Mitra and Craswell(2015)]%
        {mitra_query_2015}
\bibfield{author}{\bibinfo{person}{Bhaskar Mitra} {and} \bibinfo{person}{Nick
  Craswell}.} \bibinfo{year}{2015}\natexlab{}.
\newblock \showarticletitle{Query {Auto}-{Completion} for {Rare} {Prefixes}}.
  In \bibinfo{booktitle}{\emph{Proceedings of the 24th {ACM} {International} on
  {Conference} on {Information} and {Knowledge} {Management}}}
  \emph{(\bibinfo{series}{{CIKM} '15})}. \bibinfo{publisher}{ACM Press},
  \bibinfo{address}{New York, NY, USA}, \bibinfo{pages}{1755--1758}.
\newblock
\showISBNx{978-1-4503-3794-6}
\urldef\tempurl%
\url{https://doi.org/10.1145/2806416.2806599}
\showDOI{\tempurl}


\bibitem[Mitra et~al\mbox{.}(2014)]%
        {mitra_user_2014}
\bibfield{author}{\bibinfo{person}{Bhaskar Mitra}, \bibinfo{person}{Milad
  Shokouhi}, \bibinfo{person}{Filip Radlinski}, {and} \bibinfo{person}{Katja
  Hofmann}.} \bibinfo{year}{2014}\natexlab{}.
\newblock \showarticletitle{On user interactions with query auto-completion}.
  In \bibinfo{booktitle}{\emph{Proceedings of the 37th international {ACM}
  {SIGIR} conference on {Research} \& development in information retrieval}}
  \emph{(\bibinfo{series}{{SIGIR} '14})}. \bibinfo{publisher}{ACM Press},
  \bibinfo{address}{New York, NY, USA}, \bibinfo{pages}{1055--1058}.
\newblock
\showISBNx{978-1-4503-2257-7}
\urldef\tempurl%
\url{https://doi.org/10.1145/2600428.2609508}
\showDOI{\tempurl}


\bibitem[Nazeri et~al\mbox{.}(2019)]%
        {nazeri_edgeconnect_2019}
\bibfield{author}{\bibinfo{person}{Kamyar Nazeri}, \bibinfo{person}{Eric Ng},
  \bibinfo{person}{Tony Joseph}, \bibinfo{person}{Faisal Qureshi}, {and}
  \bibinfo{person}{Mehran Ebrahimi}.} \bibinfo{year}{2019}\natexlab{}.
\newblock \showarticletitle{{EdgeConnect}: {Structure} {Guided} {Image}
  {Inpainting} using {Edge} {Prediction}}. In \bibinfo{booktitle}{\emph{2019
  {IEEE}/{CVF} {International} {Conference} on {Computer} {Vision} {Workshop}
  ({ICCVW})}}. \bibinfo{publisher}{IEEE}, \bibinfo{address}{Seoul, Korea
  (South)}, \bibinfo{pages}{3265--3274}.
\newblock
\showISBNx{978-1-72815-023-9}
\urldef\tempurl%
\url{https://doi.org/10.1109/ICCVW.2019.00408}
\showDOI{\tempurl}


\bibitem[Pandian and Suleri(2020)]%
        {pandian_blackbox_2020}
\bibfield{author}{\bibinfo{person}{Vinoth Pandian} {and} \bibinfo{person}{Sarah
  Suleri}.} \bibinfo{year}{2020}\natexlab{}.
\newblock \showarticletitle{{BlackBox} {Toolkit}: {Intelligent} {Assistance} to
  {UI} {Design}}. In \bibinfo{booktitle}{\emph{{CHI}’20, {Workshop} on
  {Artificial} {Intelligence} for {HCI}: {A} {Modern} {Approach}}}.
\newblock


\bibitem[Park and Chiba(2017)]%
        {park_neural_2017}
\bibfield{author}{\bibinfo{person}{Dae~Hoon Park} {and} \bibinfo{person}{Rikio
  Chiba}.} \bibinfo{year}{2017}\natexlab{}.
\newblock \showarticletitle{A {Neural} {Language} {Model} for {Query}
  {Auto}-{Completion}}. In \bibinfo{booktitle}{\emph{Proceedings of the 40th
  {International} {ACM} {SIGIR} {Conference} on {Research} and {Development} in
  {Information} {Retrieval}}} \emph{(\bibinfo{series}{{SIGIR} '17})}.
  \bibinfo{publisher}{ACM Press}, \bibinfo{address}{New York, NY, USA},
  \bibinfo{pages}{1189--1192}.
\newblock
\showISBNx{978-1-4503-5022-8}
\urldef\tempurl%
\url{https://doi.org/10.1145/3077136.3080758}
\showDOI{\tempurl}


\bibitem[Quinn and Zhai(2016)]%
        {quinn_cost-benefit_2016}
\bibfield{author}{\bibinfo{person}{Philip Quinn} {and} \bibinfo{person}{Shumin
  Zhai}.} \bibinfo{year}{2016}\natexlab{}.
\newblock \showarticletitle{A {Cost}-{Benefit} {Study} of {Text} {Entry}
  {Suggestion} {Interaction}}. In \bibinfo{booktitle}{\emph{Proceedings of the
  2016 {CHI} {Conference} on {Human} {Factors} in {Computing} {Systems}}}.
  \bibinfo{publisher}{ACM Press}, \bibinfo{address}{New York, NY, USA},
  \bibinfo{pages}{83--88}.
\newblock
\showISBNx{978-1-4503-3362-7}
\urldef\tempurl%
\url{https://doi.org/10.1145/2858036.2858305}
\showDOI{\tempurl}


\bibitem[Radford et~al\mbox{.}(2019)]%
        {radford_language_2019}
\bibfield{author}{\bibinfo{person}{Alec Radford}, \bibinfo{person}{Jeffrey Wu},
  \bibinfo{person}{Rewon Child}, \bibinfo{person}{David Luan},
  \bibinfo{person}{Dario Amodei}, {and} \bibinfo{person}{Ilya Sutskever}.}
  \bibinfo{year}{2019}\natexlab{}.
\newblock \showarticletitle{Language {Models} are {Unsupervised} {Multitask}
  {Learners}}.
\newblock  (\bibinfo{year}{2019}), \bibinfo{pages}{24}.
\newblock
\urldef\tempurl%
\url{https://cdn.openai.com/better-language-models/language_models_are_unsupervised_multitask_learners.pdf}
\showURL{%
\tempurl}


\bibitem[Shokouhi(2013)]%
        {shokouhi_learning_2013}
\bibfield{author}{\bibinfo{person}{Milad Shokouhi}.}
  \bibinfo{year}{2013}\natexlab{}.
\newblock \showarticletitle{Learning to personalize query auto-completion}. In
  \bibinfo{booktitle}{\emph{Proceedings of the 36th international {ACM} {SIGIR}
  conference on {Research} and development in information retrieval}}
  \emph{(\bibinfo{series}{{SIGIR} '13})}. \bibinfo{publisher}{ACM Press},
  \bibinfo{address}{New York, NY, USA}, \bibinfo{pages}{103--112}.
\newblock
\showISBNx{978-1-4503-2034-4}
\urldef\tempurl%
\url{https://doi.org/10.1145/2484028.2484076}
\showDOI{\tempurl}


\bibitem[Shokouhi and Radinsky(2012)]%
        {shokouhi_time-sensitive_2012}
\bibfield{author}{\bibinfo{person}{Milad Shokouhi} {and} \bibinfo{person}{Kira
  Radinsky}.} \bibinfo{year}{2012}\natexlab{}.
\newblock \showarticletitle{Time-sensitive query auto-completion}. In
  \bibinfo{booktitle}{\emph{Proceedings of the 35th international {ACM} {SIGIR}
  conference on {Research} and development in information retrieval}}
  \emph{(\bibinfo{series}{{SIGIR} '12})}. \bibinfo{publisher}{ACM Press},
  \bibinfo{address}{New York, NY, USA}, \bibinfo{pages}{601--610}.
\newblock
\showISBNx{978-1-4503-1472-5}
\urldef\tempurl%
\url{https://doi.org/10.1145/2348283.2348364}
\showDOI{\tempurl}


\bibitem[Svyatkovskiy et~al\mbox{.}(2019)]%
        {svyatkovskiy_pythia_2019}
\bibfield{author}{\bibinfo{person}{Alexey Svyatkovskiy}, \bibinfo{person}{Ying
  Zhao}, \bibinfo{person}{Shengyu Fu}, {and} \bibinfo{person}{Neel
  Sundaresan}.} \bibinfo{year}{2019}\natexlab{}.
\newblock \showarticletitle{Pythia: {AI}-assisted {Code} {Completion}
  {System}}. In \bibinfo{booktitle}{\emph{{KDD} '19: {Proceedings} of the 25th
  {ACM} {SIGKDD} {International} {Conference} on {Knowledge} {Discovery} \&
  {Data} {Mining}}} \emph{(\bibinfo{series}{{KDD} '19})}.
  \bibinfo{publisher}{ACM Press}, \bibinfo{address}{New York, NY, USA},
  \bibinfo{pages}{2727--2735}.
\newblock
\showISBNx{978-1-4503-6201-6}
\urldef\tempurl%
\url{https://doi.org/10.1145/3292500.3330699}
\showDOI{\tempurl}


\bibitem[Tirkaz et~al\mbox{.}(2012)]%
        {tirkaz_sketched_2012}
\bibfield{author}{\bibinfo{person}{Caglar Tirkaz}, \bibinfo{person}{Berrin
  Yanikoglu}, {and} \bibinfo{person}{T. Metin~Sezgin}.}
  \bibinfo{year}{2012}\natexlab{}.
\newblock \showarticletitle{Sketched symbol recognition with auto-completion}.
\newblock \bibinfo{journal}{\emph{Pattern Recognition}} \bibinfo{volume}{45},
  \bibinfo{number}{11} (\bibinfo{date}{Nov.} \bibinfo{year}{2012}),
  \bibinfo{pages}{3926--3937}.
\newblock
\showISSN{0031-3203}
\urldef\tempurl%
\url{https://doi.org/10.1016/j.patcog.2012.04.026}
\showDOI{\tempurl}


\bibitem[Vaswani et~al\mbox{.}(2017)]%
        {vaswani_attention_2017}
\bibfield{author}{\bibinfo{person}{Ashish Vaswani}, \bibinfo{person}{Noam
  Shazeer}, \bibinfo{person}{Niki Parmar}, \bibinfo{person}{Jakob Uszkoreit},
  \bibinfo{person}{Llion Jones}, \bibinfo{person}{Aidan~N. Gomez},
  \bibinfo{person}{Łukasz Kaiser}, {and} \bibinfo{person}{Illia Polosukhin}.}
  \bibinfo{year}{2017}\natexlab{}.
\newblock \showarticletitle{Attention is all you need}. In
  \bibinfo{booktitle}{\emph{Proceedings of the 31st {International}
  {Conference} on {Neural} {Information} {Processing} {Systems}}}
  \emph{(\bibinfo{series}{{NIPS}'17})}. \bibinfo{publisher}{Curran Associates
  Inc.}, \bibinfo{address}{Long Beach, California, USA},
  \bibinfo{pages}{6000--6010}.
\newblock
\showISBNx{978-1-5108-6096-4}


\bibitem[Wang et~al\mbox{.}(2018)]%
        {wang_image_2018}
\bibfield{author}{\bibinfo{person}{Yi Wang}, \bibinfo{person}{Xin Tao},
  \bibinfo{person}{Xiaojuan Qi}, \bibinfo{person}{Xiaoyong Shen}, {and}
  \bibinfo{person}{Jiaya Jia}.} \bibinfo{year}{2018}\natexlab{}.
\newblock \showarticletitle{Image {Inpainting} via {Generative} {Multi}-column
  {Convolutional} {Neural} {Networks}}. In \bibinfo{booktitle}{\emph{Advances
  in {Neural} {Information} {Processing} {Systems} 31: {Annual} {Conference} on
  {Neural} {Information} {Processing} {Systems} 2018}}.
  \bibinfo{publisher}{Curran Associates Inc.}
\newblock
\urldef\tempurl%
\url{https://doi.org/10.5555/3326943.3326974}
\showDOI{\tempurl}
\newblock
\shownote{arXiv: 1810.08771}.


\bibitem[Yang et~al\mbox{.}(2018)]%
        {yang_mapping_2018}
\bibfield{author}{\bibinfo{person}{Qian Yang}, \bibinfo{person}{Nikola
  Banovic}, {and} \bibinfo{person}{John Zimmerman}.}
  \bibinfo{year}{2018}\natexlab{}.
\newblock \showarticletitle{Mapping {Machine} {Learning} {Advances} from {HCI}
  {Research} to {Reveal} {Starting} {Places} for {Design} {Innovation}}. In
  \bibinfo{booktitle}{\emph{Proceedings of the 2018 {CHI} {Conference} on
  {Human} {Factors} in {Computing} {Systems} - {CHI} '18}}.
  \bibinfo{publisher}{ACM Press}, \bibinfo{address}{New York, NY, USA},
  \bibinfo{pages}{1--11}.
\newblock
\showISBNx{978-1-4503-5620-6}
\urldef\tempurl%
\url{https://doi.org/10.1145/3173574.3173704}
\showDOI{\tempurl}


\bibitem[Yang et~al\mbox{.}(2020)]%
        {yang_re-examining_2020}
\bibfield{author}{\bibinfo{person}{Qian Yang}, \bibinfo{person}{Aaron
  Steinfeld}, \bibinfo{person}{Carolyn Rosé}, {and} \bibinfo{person}{John
  Zimmerman}.} \bibinfo{year}{2020}\natexlab{}.
\newblock \showarticletitle{Re-examining {Whether}, {Why}, and {How}
  {Human}-{AI} {Interaction} {Is} {Uniquely} {Difficult} to {Design}}. In
  \bibinfo{booktitle}{\emph{Proceedings of the 2020 {CHI} {Conference} on
  {Human} {Factors} in {Computing} {Systems}}} \emph{(\bibinfo{series}{{CHI}
  '20})}. \bibinfo{publisher}{ACM Press}, \bibinfo{address}{New York, NY, USA},
  \bibinfo{pages}{1--13}.
\newblock
\showISBNx{978-1-4503-6708-0}
\urldef\tempurl%
\url{https://doi.org/10.1145/3313831.3376301}
\showDOI{\tempurl}


\bibitem[Yang et~al\mbox{.}(2016)]%
        {yang_planning_2016}
\bibfield{author}{\bibinfo{person}{Qian Yang}, \bibinfo{person}{John
  Zimmerman}, \bibinfo{person}{Aaron Steinfeld}, {and} \bibinfo{person}{Anthony
  Tomasic}.} \bibinfo{year}{2016}\natexlab{}.
\newblock \showarticletitle{Planning {Adaptive} {Mobile} {Experiences} {When}
  {Wireframing}}. In \bibinfo{booktitle}{\emph{Proceedings of the 2016 {ACM}
  {Conference} on {Designing} {Interactive} {Systems} - {DIS} '16}}.
  \bibinfo{publisher}{ACM Press}, \bibinfo{address}{New York, NY, USA},
  \bibinfo{pages}{565--576}.
\newblock
\showISBNx{978-1-4503-4031-1}
\urldef\tempurl%
\url{https://doi.org/10.1145/2901790.2901858}
\showDOI{\tempurl}


\bibitem[Yannakakis et~al\mbox{.}(2014)]%
        {yannakakis_mixed-initiative_2014}
\bibfield{author}{\bibinfo{person}{Georgios~N Yannakakis},
  \bibinfo{person}{Antonios Liapis}, {and} \bibinfo{person}{Constantine
  Alexopoulos}.} \bibinfo{year}{2014}\natexlab{}.
\newblock \showarticletitle{Mixed-initiative co-creativity}.
\newblock \bibinfo{journal}{\emph{9th International Conference on the
  Foundations of Digital Games}} (\bibinfo{year}{2014}), \bibinfo{pages}{8}.
\newblock
\urldef\tempurl%
\url{https://www.um.edu.mt/library/oar//handle/123456789/29459}
\showURL{%
\tempurl}


\bibitem[Yu et~al\mbox{.}(2018)]%
        {yu_generative_2018}
\bibfield{author}{\bibinfo{person}{Jiahui Yu}, \bibinfo{person}{Zhe Lin},
  \bibinfo{person}{Jimei Yang}, \bibinfo{person}{Xiaohui Shen},
  \bibinfo{person}{Xin Lu}, {and} \bibinfo{person}{Thomas~S Huang}.}
  \bibinfo{year}{2018}\natexlab{}.
\newblock \showarticletitle{Generative {Image} {Inpainting} with {Contextual}
  {Attention}}. In \bibinfo{booktitle}{\emph{2018 {IEEE}/{CVF} {Conference} on
  {Computer} {Vision} and {Pattern} {Recognition}}}. \bibinfo{address}{Salt
  Lake City, UT, USA}, \bibinfo{pages}{5505--5514}.
\newblock
\urldef\tempurl%
\url{https://doi.org/10.1109/CVPR.2018.00577}
\showDOI{\tempurl}


\bibitem[Zhang et~al\mbox{.}(2016)]%
        {zhang_towards_2016}
\bibfield{author}{\bibinfo{person}{Aston Zhang}, \bibinfo{person}{Amit Goyal},
  \bibinfo{person}{Ricardo Baeza-Yates}, \bibinfo{person}{Yi Chang},
  \bibinfo{person}{Jiawei Han}, \bibinfo{person}{Carl~A. Gunter}, {and}
  \bibinfo{person}{Hongbo Deng}.} \bibinfo{year}{2016}\natexlab{}.
\newblock \showarticletitle{Towards {Mobile} {Query} {Auto}-{Completion}: {An}
  {Efficient} {Mobile} {Application}-{Aware} {Approach}}. In
  \bibinfo{booktitle}{\emph{Proceedings of the 25th {International}
  {Conference} on {World} {Wide} {Web}}} \emph{(\bibinfo{series}{{WWW} '16})}.
  \bibinfo{publisher}{International World Wide Web Conferences Steering
  Committee}, \bibinfo{address}{Montréal, Québec, Canada},
  \bibinfo{pages}{579--590}.
\newblock
\showISBNx{978-1-4503-4143-1}
\urldef\tempurl%
\url{https://doi.org/10.1145/2872427.2882977}
\showDOI{\tempurl}


\bibitem[Zhang et~al\mbox{.}(2015)]%
        {zhang_adaqac_2015}
\bibfield{author}{\bibinfo{person}{Aston Zhang}, \bibinfo{person}{Amit Goyal},
  \bibinfo{person}{Weize Kong}, \bibinfo{person}{Hongbo Deng},
  \bibinfo{person}{Anlei Dong}, \bibinfo{person}{Yi Chang},
  \bibinfo{person}{Carl~A. Gunter}, {and} \bibinfo{person}{Jiawei Han}.}
  \bibinfo{year}{2015}\natexlab{}.
\newblock \showarticletitle{{adaQAC}: {Adaptive} {Query} {Auto}-{Completion}
  via {Implicit} {Negative} {Feedback}}. In
  \bibinfo{booktitle}{\emph{Proceedings of the 38th {International} {ACM}
  {SIGIR} {Conference} on {Research} and {Development} in {Information}
  {Retrieval}}} \emph{(\bibinfo{series}{{SIGIR} '15})}. \bibinfo{publisher}{ACM
  Press}, \bibinfo{address}{New York, NY, USA}, \bibinfo{pages}{143--152}.
\newblock
\showISBNx{978-1-4503-3621-5}
\urldef\tempurl%
\url{https://doi.org/10.1145/2766462.2767697}
\showDOI{\tempurl}


\bibitem[Zhang and Balog(2019)]%
        {zhang_auto-completion_2019}
\bibfield{author}{\bibinfo{person}{Shuo Zhang} {and} \bibinfo{person}{Krisztian
  Balog}.} \bibinfo{year}{2019}\natexlab{}.
\newblock \showarticletitle{Auto-completion for {Data} {Cells} in {Relational}
  {Tables}}. In \bibinfo{booktitle}{\emph{Proceedings of the 28th {ACM}
  {International} {Conference} on {Information} and {Knowledge} {Management}}}
  \emph{(\bibinfo{series}{{CIKM} '19})}. \bibinfo{publisher}{ACM Press},
  \bibinfo{address}{New York, NY, USA}, \bibinfo{pages}{761--770}.
\newblock
\showISBNx{978-1-4503-6976-3}
\urldef\tempurl%
\url{https://doi.org/10.1145/3357384.3357932}
\showDOI{\tempurl}


\bibitem[Zhang and van~de Panne(2018)]%
        {zhang_data-driven_2018}
\bibfield{author}{\bibinfo{person}{Xinyi Zhang} {and} \bibinfo{person}{Michiel
  van~de Panne}.} \bibinfo{year}{2018}\natexlab{}.
\newblock \showarticletitle{Data-driven autocompletion for keyframe animation}.
  In \bibinfo{booktitle}{\emph{Proceedings of the 11th {Annual} {International}
  {Conference} on {Motion}, {Interaction}, and {Games}}}
  \emph{(\bibinfo{series}{{MIG} '18})}. \bibinfo{publisher}{ACM Press},
  \bibinfo{address}{New York, NY, USA}, \bibinfo{pages}{1--11}.
\newblock
\showISBNx{978-1-4503-6015-9}
\urldef\tempurl%
\url{https://doi.org/10.1145/3274247.3274502}
\showDOI{\tempurl}


\end{thebibliography}

\end{document}